# Strongly enhancend anisotropic upper critical field in carbon substituted MgB$_2$: impact of Fermi surface changes

R. Puzniak,[1, a] M. Angst,[2, 3] A. Szewczyk,[1] J. Jun,[4] S. M. Kazakov,[4] and J. Karpinski[4]

[1] Institute of Physics, Polish Academy of Sciences, Aleja Lotników 32/46, PL 02-668 Warszawa, Poland

[2] Physik-Institut der Universität Zürich, CH-8057 Zürich, Switzerland

[3] Ames Laboratory, Ames, IA 50011, USA

[4] Solid State Physics Laboratory, ETH Zürich, CH-8093 Zürich, Switzerland

The upper critical field $H_{c2}^{\|c}(0) \cong 85$ kOe in Mg(B$_{0.94}$C$_{0.06}$)$_2$ single crystals, determined from torque measurements, is more than twice as large as that one of $\cong 31$ kOe in unsubstituted MgB$_2$. Anisotropy of $H_{c2}$ increases from $\gamma_H \approx 3.4$ near $T_c$, the value close to that of $\gamma_H(T_c)$ in MgB$_2$, to about 4 at low temperature, the value considerably lower than that in MgB$_2$. The corresponding $H_{c2}^{\|ab}(0) \approx 330$-$350$ kOe is likely close to the maximum enhancement due to C substitution. The position of a sharp peak in the irreversible torque $H_{max}$ was found to be shifted to lower reduced fields $H/H_{c2}$ in comparison to its position in unsubstituted MgB$_2$, indicating increasing disorder. The enhancement of $H_{c2}$ can be explained as a disorder effect only if the main result of disorder is to make the $\pi$ bands more dirty while not affecting the $\sigma$ bands as much. In addition to disorder and weakened electron-phonon coupling, the impact of the Fermi level shifting into a region with lower $\sigma$ Fermi velocities has to be taken into account in the analysis of $H_{c2}$ data as well.

74.60.Ec, 74.25.Dw, 74.25.Ha, 74.70.Ad

---

[a] Email address: puzni@ifpan.edu.pl (R. Puzniak)

Superconducting MgB$_2$ (Ref. 1) exhibits rather peculiar properties,[2] originating from the involvement of two sets of bands of different anisotropy and different coupling to the phonons.[3] Among them are pronounced deviations of the upper critical field, $H_{c2}$, from predictions of the widely used anisotropic Ginzburg-Landau theory (AGLT).[4, 5, 6, 7, 8, 9, 10, 11] MgB$_2$ is considered also to have great potential for applications because of its relatively high transition temperature $T_c \approx 39$ and sufficiently large coherence length to overcome the problem of weak links limiting any possible large scale applications of high $T_c$ superconductors.[12] Problematic is that the out-of-plane upper critical field $H_{c2}^{\|c}(0) \cong 31$ kOe (Ref. 4) is low. Substitutional chemistry is one of the most effective methods to modify the superconducting properties of a material.[13] Partial replacement of boron by carbon appears to be one of the most interesting routes, leading to the material with reduced $T_c$, but considerable enhancements of $H_{c2}$ and reduced anisotropy.[14, 15, 16, 17, 18, 19] Specific heat data suggest that two-gap superconductivity is preserved in the Mg(B$_{0.9}$C$_{0.1}$)$_2$ despite the heavily suppressed $T_c$.[14] Spectroscopic measurements support this, finding reduced values of the smaller energy gap on the π bands, $2\Delta_\pi/k_B T_c$, in C substituted materials to be the same as in MgB$_2$.[20]

Here, we present the results of torque measurements performed on a single crystal of Mg(B$_{0.94}$C$_{0.06}$)$_2$. A significant enhancement of upper critical field, a much weaker temperature dependence of its anisotropy and a large shift of the position of a sharp peak effect (PE) in the irreversible torque to lower reduced fields as an effect of carbon doping MgB$_2$ were found. While the downward shift of the PE is most likely due to the increased disorder resulting from the partial substitution, we will demonstrate that neither the enhancement of $H_{c2}$ nor the decrease of its anisotropy can be explained by simple disorder effects only, and the impact of the electron doping by the C substitution has to be taken into account in the analysis of $H_{c2}$ data as well.

Single crystals of $Mg(B_{0.94}C_{0.06})_2$ were grown in a cubic anvil system at high pressure/temperature conditions similar to those described in Ref. 21. Magnesium (Fluka, >99% purity), amorphous boron (Alfa Aesar, >99.99%), and graphite powder (Alfa Aesar, >99.99%) were used as starting materials. The boron was annealed under dynamic vacuum at 1200°C to remove moisture. A mixture of magnesium and boron, with a 5% substitution of boron by graphite, was pressed into a pellet and put into a BN container of 6 mm internal diameter and 8 mm length. The pressure was applied at room temperature in the system with a pyrophylite cube as a medium and then the temperature increased during one hour, up to a maximum of 1900-1950°C, kept for 30 min, and decreased during 1-2 hours. Flat crystals were grown with dimensions up to $0.8\times0.8\times0.2$ mm$^3$ and sharp superconducting transitions at 33-33.5 K. They were black in color in contrast to golden non-substituted $MgB_2$. The carbon content of the crystals, $x \approx 0.063$, was estimated from the $a$ lattice parameter according to the data of Ref. 22, assuming a linear dependence of $a$ on the carbon content. Although this is a rough approximation, no other proof of the carbon content was found and the refinement of x-ray data was impossible because of similar scattering factors for B and C atoms.[23] A non-linear increase of the normalized resistance $R(40\ K)/R(300\ K)$ $1/rrr$, where $rrr$ is residual resistance ratio, from about 0.15 for unsubstituted $MgB_2$ to 0.5 for $Mg(B_{0.95}C_{0.05})_2$ and 0.7 for $Mg(B_{0.905}C_{0.095})_2$, was observed.[23] The changes in the residual resistance ratio may indicate that the importance of scattering by impurities is increased compared to scattering by phonons in C substituted crystals.[24] Direct comparison of the in-plane residual resistivity of C subsituted $Mg(B_{0.95}C_{0.05})_2$ with $\rho_{ab}(40\ K) = 10$ μΩcm (Ref. 23) to the one of unsubstituted $MgB_2$ with $\rho_{ab}(40\ K) = 2$ μΩcm (Ref. 25) clearly shows that, we have a 5-fold increase of the residual resistivity of 5%C and 0%C crystals from same source – definitely relevant and suggesting increased impurity scattering.

A high-quality single crystal of Mg(B$_{0.94}$C$_{0.06}$)$_2$ with a volume of about $6\times10^{-3}$ mm$^3$ was selected for torque measurements. The $T$ dependence of the magnetization of the crystal in 3 Oe [right inset of Fig. 1a)] shows at 33.45 K a sharp transition to the superconducting state, comparable with the one of the unsubstituted crystals studied in Refs. 4 and 26. Measurements to study the upper critical field and the peak effect were carried out in magnetic fields up to 90 kOe and temperatures down to 2.2 K with the torque option of a Quantum Design PPMS (physical property measurement system). The torque $\boldsymbol{\tau} = \boldsymbol{m} \times \boldsymbol{B} \cong \boldsymbol{m} \times \boldsymbol{H}$, where $\boldsymbol{m}$ is the magnetic moment of the sample, was recorded as a function of the angle $\theta$ between the applied field $\boldsymbol{H}$ and the $c$ axis of the crystal for various fixed temperatures and fields.

An example of a torque vs angle curve recorded in 50 kOe at a 19 K is given in Fig. 1a)[▽]. For small angles $\theta$ the torque is essentially zero. Only for $\theta > 41.5$ deg is there an appreciable torque signal. The crossover angle $\theta_{c2}$ between the normal and the superconducting state is the angle for which the fixed applied field is the upper critical field. For the unsubstituted MgB$_2$ at the same $T/T_c \cong 0.57$, the crossover angle is 26.2 deg in 19 kOe (left inset) and to 45 deg in 24 kOe.[4] The absence of an angular region with zero torque for Mg(B$_{0.94}$C$_{0.06}$)$_2$ at $0.57T_c$ in 24 kOe [■] implies that for this compound $H_{c2}$ for all field orientations is higher than 24 kOe. The clear difference in the $H$ dependence of $\theta_{c2}$ between Mg(B$_{0.94}$C$_{0.06}$)$_2$ and MgB$_2$ at the same reduced temperature provides direct evidence of a more than double increase of upper critical field as a result of carbon substitution into the boron site of MgB$_2$. The increase of $H_{c2}$ is accompanied by a sharp peak in the irreversible torque at the fields/angles close to the upper critical field/angle [▽]. The peak develops with increasing $H$ and decreasing $T$ [see Fig. 1b)]. The peak amplitude, comparable to or even smaller than the reversible torque 50 kOe at $0.31T_c$, expands in 70 kOe to values almost an order of magnitude higher than the reversible torque.

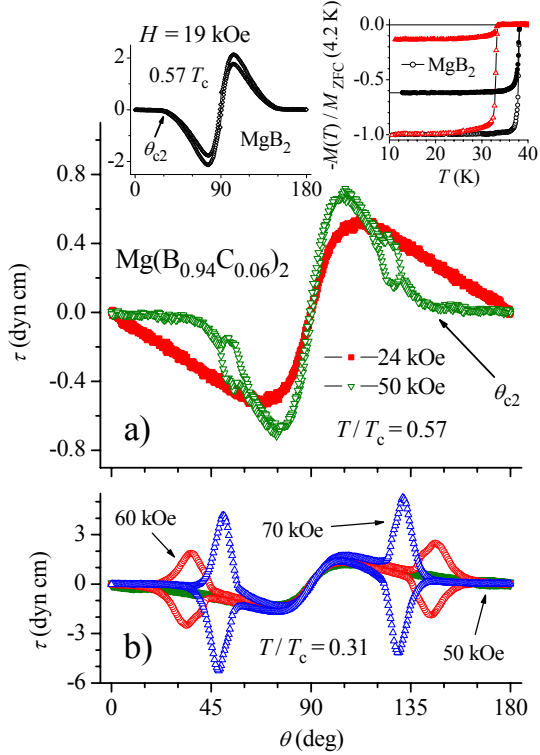

**FIG. 1. Torque $\tau$ versus angle $\theta$ between the applied field and the *c*-axis of Mg(B$_{0.94}$C$_{0.06}$)$_2$ single crystal, at $0.57T_c$ in magnetic fields of 24 and 50 kOe [a)] and at $0.31T_c$ in a field of 50, 60, and 70 kOe [b)]. $\theta_{c2}$ marks the crossover angle between normal and superconducting state. The value of $\theta_{c2}$ for Mg(B$_{0.94}$C$_{0.06}$)$_2$ at $0.57T_c$ in 50 kOe is very similar to the one for MgB$_2$ at the same reduced temperature in 19 kOe [left inset of panel a), data from Ref. 27]. a) right inset: zero field (open symbols) and field cooled (full symbols) $M(T)$ dependence for Mg(B$_{0.94}$C$_{0.06}$)$_2$ in 3 Oe (triangles) and for MgB$_2$ in 2 Oe (circles, Ref. 28), $H\|c$.**

To determine $H_{c2}$ more accurately we applied the appropriate scaling analysis as described in Ref. 4. The universal rescaled torque signal function of the distance from $T_c$ with a fixed value at $T = T_c(H)$ was found to be[4] $P = -\tau \varepsilon^{1/3}(\theta) / \left( \sin\theta \cos\theta \, H^{5/3} (1 - 1/\gamma_H^2) T^{2/3} \right)$, with $\varepsilon(\theta) = (\cos^2\theta + \sin^2\theta/\gamma_H^2)^{1/2}$ and $\gamma_H = H_{c2}^{\|ab}/H_{c2}^{\|c}$ determined self-consistently. In the present case of a sample volume of about $6\times10^{-3}$ mm$^3$, the proper criterion for $H_{c2}$ is P = $1.6\times10^{-10}$ dyn cm Oe$^{-5/3}$K$^{-2/3}$. It was already shown[4] that variations of the criterion up to a factor of 3 (to take into account uncertainties in the sample volume determination) yield very similar $H_{c2}(T)$ and $\gamma_H(T)$. The so determined $H_{c2}(\theta)$ at $0.57T_c$ is shown in Fig. 2.

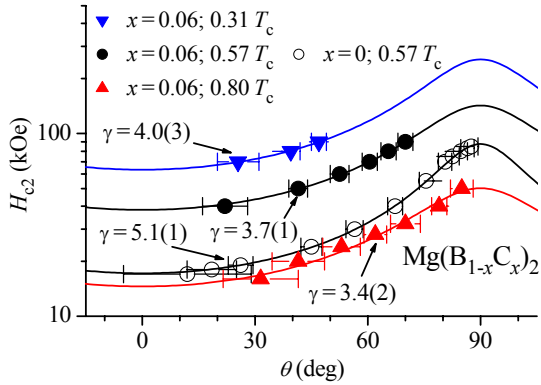

**FIG. 2. Upper critical field $H_{c2}$ vs $T$ for Mg(B$_{0.94}$C$_{0.06}$)$_2$ single crystal (full symbols) at $0.31T_c$, $0.57T_c$, and $0.80T_c$ and for MgB$_2$ crystal (open symbols) at $0.57T_c$ (circles, data from Ref. 4). The full lines present a free fit of Eq. (1) to the data.**

Within AGLT the angle dependence of the upper critical is predicted to be[29]

$$H_{c2}(\theta) = H_{c2}^{\|c}\left(\cos^2\theta + \sin^2\theta/\gamma_H^2\right)^{-1/2}. \qquad (1)$$

A fit of Eq. (1) to the data at 19 K yields $\gamma_H = 3.7(1)$ and $H_{c2}^{\|c} = 38.2$ kOe. The upper critical field parallel to the $c$-axis of Mg(B$_{0.94}$C$_{0.06}$)$_2$ is more than twice as high as the one of 17.2 kOe obtained for MgB$_2$ at the same $T/T_{c2} \cong 0.57$, whereas the anisotropy $\gamma_H$ is 27% lower. Since the $H_{c2}(\theta)$ dependences for Mg(B$_{0.94}$C$_{0.06}$)$_2$ and for MgB$_2$ at the same reduced temperature are not two parallel curves with a logarithmic scale in $H$, the difference in the anisotropy of $H_{c2}$ between the two compounds is as obvious as the difference in the magnitude. Before discussing the temperature dependence of $H_{c2}$ and $\gamma_H$ we turn to the details of the irreversible torque behavior.

The peak in the irreversible part of torque dependence [Fig. 1b)] is well pronounced and very sharp for fields of 60 and 70 kOe. Two characteristic fields may be easily determined: the field of the maximum of torque difference between the values for clockwise ($\tau_{inc}$) and counterclockwise ($\tau_{dec}$) change of the field direction, $H_{max}$, and the irreversibility field $H_{irr}$, where the two branches of the hysteresis loop meet. The $\theta$ dependence of these and $H_{c2}$ are compared to the ones for unsubstituted MgB$_2$ at the same $T/T_c \cong 0.57$ in Fig. 3. While the characteristic peak fields for MgB$_2$ are located very close to $H_{c2}$ and follow its angular dependence, the angular scaling of $H_{max}$ is much less clear for Mg(B$_{0.94}$C$_{0.06}$)$_2$, which is emphasized in the left inset. For MgB$_2$, $H_{max} \cong 0.85 H_{c2}$ independent of the reduced temperature.[26] For Mg(B$_{0.94}$C$_{0.06}$)$_2$ the ratio between $H_{max}$ and $H_{c2}$ at $T/T_c \cong 0.57$ and 30 deg does not differ much from the one for MgB$_2$, but $H_{max}/H_{c2}$ decreases with increasing $\theta$. The effect is much more pronounced at lower $T$.

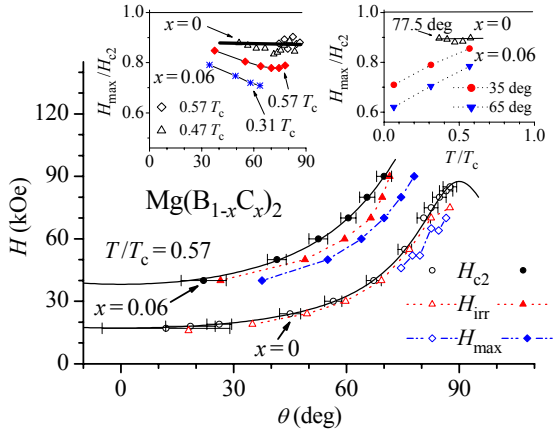

**Fig. 3. Angle dependence of upper critical field $H_{c2}$ (circles), the irreversibility field $H_{irr}$ (triangles), and the peak maximum field $H_{max}$ (diamonds) at $0.57T_c$ for Mg(B$_{0.94}$C$_{0.06}$)$_2$ (full symbols) and for MgB$_2$ (open symbols). Full lines are fits of the theoretical $H_{c2}(\theta)$ dependence [Eq. (1)] Dashed lines are guides for eye. Left inset: $\theta$ dependence of the**

**reduced peak maximum field for Mg(B$_{0.94}$C$_{0.06}$)$_2$ at 0.31$T_c$ (✶) and 0.57$T_c$ (◆), compared with the one for MgB$_2$ at 0.47$T_c$ (△) and 0.57$T_c$ (◇). Right inset: reduced peak maximum field vs $T/T_c$ for Mg(B$_{0.94}$C$_{0.06}$)$_2$ at 35 (●) and 65 deg (▼) between the applied field and the $c$-axis of the crystal compared with that one for MgB$_2$ at 77.5 deg (△).**

In a Ref. 26, we demonstrated that the peak effect (PE) signifies a disorder-induced phase transition of vortex matter. Briefly examined alternative origins of the PE due to inhomogeneities or extended defects or due to change of the elastic constants of the vortex lattice when $H$ approaches $H_{c2}$, not associated with a phase transition, were excluded. Since the arguments against alternative origins of the PE in MgB$_2$ are also valid for the carbon substituted compound, the assumption of the same PE origin in Mg(B$_{0.94}$C$_{0.06}$)$_2$ seems reasonable. The considerably lower reduced field position of the PE at low temperature in Mg(B$_{0.94}$C$_{0.06}$)$_2$ indicates, then, that the amount of random pointlike disorder is increased by the C substitution (cf. Ref. 8), as may be expected. However, remarkable is the pronounced deviation of the angular scaling of $H_{max}$ from the one of $H_{c2}$ in Mg(B$_{0.94}$C$_{0.06}$)$_2$, whereas there is only a weak indication for such a deviation in MgB$_2$ (see Fig. 3 in Ref. 26). Since the order-disorder transition is determined by the competition of the elastic energy $E_{el}$ of the vortex lattice and the pinning energy $E_{pin}$ due to the pointlike disorder, an angular dependence of $H_{max}/H_{c2}$ may arise from anisotropic pinning and/or deviations of the vortex lattice energies from AGLT. To explain the observed $\theta$ dependence of $H_{max}/H_{c2}$, anisotropic pinning would have to be associated with stronger impurity scattering perpendicular to the planes, arising, e.g., from a non-random modulation of C content perpendicular to the layers.[30]

The upper critical field data of Mg(B$_{0.94}$C$_{0.06}$)$_2$ are summarized in Fig. 4, compared to the data obtained previously[4] for MgB$_2$. The $T$ dependence of $H_{c2}^{\|c}$ of Mg(B$_{0.94}$C$_{0.06}$)$_2$ is in rough agreement with $H_{c2}^{\|c}(T)$ of MgB$_2$ although the downward curvature is slightly less

pronounced [inset of panel a)]. The corresponding $H_{c2}^{\|c}(0) \cong 85$ kOe is 2.7 times higher than the value of 31 kOe obtained for parent $MgB_2$.[4] As in the unsubstituted compounds there is a slight positive curvature of $H_{c2}^{\|ab}(T)$ [panel a)], which corresponds to the $T$ dependence of $\gamma_H$ [panel b)]. Because of the experimental field limitation of 90 kOe and the variation of $\gamma_H$ with $T$, only a rough estimation $H_{c2}^{\|ab}(0) \approx 330\text{-}350$ kOe can be given. The anisotropy systematically decreases with increasing temperature, similar to the case of $MgB_2$, but $\gamma_H$ is smaller at low $T$ and its variation with $T$ weaker than in $MgB_2$.

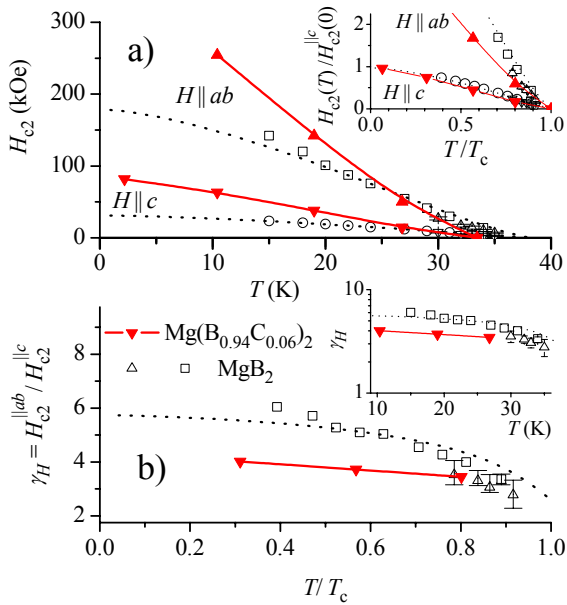

**FIG. 4. a) Upper critical field $H_{c2}$ vs temperature $T$ for $Mg(B_{0.94}C_{0.06})_2$ (full symbols) and for $MgB_2$ (open symbols, data from two different samples[4]). Triangles up and squares correspond to $H\|ab$, triangles down and circles to $H\|c$, from fits of Eq. (1) to $H_{c2}(\theta)$ data. Inset: Reduced upper critical field $H_{c2}(T)/H_{c2}^{\|c}(0)$ vs reduced temperature. b) Reduced temperature $T/T_c$ dependence of the upper critical field anisotropy $\gamma_H = H_{c2}^{\|ab}/H_{c2}^{\|c}$, determined from fits of Eq. (1) to $H_{c2}(\theta)$ for $Mg(B_{0.94}C_{0.06})_2$ and for $MgB_2$. Inset: Logarithmic dependence of $\gamma_H$ as a function of real temperature.**

Within standard one-band theory of superconductors in the dirty limit, the increase of the upper critical field upon carbon substitution might be explained by a decrease of the mean free path due to the introduction of defects. Such an explanation would be in agreement with the shift of the peak effect to lower reduced fields as well as with the increased residual resistivity. However, assuming an extrinsic origin of the upper critical field increase, in our case the decreased maximum anisotropy $\gamma_H(0)$ [Fig. 4b)] could be explained by stronger impurity scattering along the planes rising $H_{c2}^{\parallel c}$ more, incompatible with the $\theta$ dependence of $H_{max}/H_{c2}$ presented in Fig. 3. More strictly, in the case of MgB$_2$ we need to consider three different types of impurity scattering:[24] interband scattering and intraband scattering within the σ and π bands, which yield rather different mean free paths.[31] Partial carbon substitution may well affect the intraband scattering rates and effective coherence lengths in σ and π bands differently; since two-gap superconductivity survives with essentially identical $2\Delta_\pi/k_B T_c$ values, it is clear that *interband scattering is hardly increased* by the substitution. Recently, the effect of tuning independently the scattering rates in σ and π bands on $H_{c2}$ and $\gamma_H$ was studied theoretically, with an approach based on an "intraband dirty limit".[10] The theory may account for considerable increases of $H_{c2}$. However, for an anisotropy decreasing with increasing temperature, as is the case for single crystals both in pure and C doped MgB$_2$, the theory of Ref. 10 implies a higher scattering in the σ band. For unsubstituted single crystals it is not clear though that the dirty limit condition holds for both bands, since e.g. de Haas-van-Alphen measurements[32] rather indicate mean free paths in the σ bands considerable larger than the estimated coherence lengths.[33] Furthermore, for a substitution within the boron layers, it seems natural to assume increased scattering mainly in the σ bands and expect from the theory $\gamma_H$ to decrease even stronger with increasing $T$. However, we observe the *opposite* [Fig. 4b)].

The considerations so far would seem to suggest that even in our C doped crystal, the dirty limit conditions may not be fulfilled, at least not for both sets of bands, and neither the decreased anisotropy nor the strong general enhancement of the upper critical field are simple disorder effects. *Apart from* increasing the *disorder*, the main effect of partial carbon substitution is to *dope the system with additional electrons*, shifting the Fermi energy $E_F$ to higher values. Whereas such a shift of $E_F$ does not radically change the situation in the $\pi$ bands, a not too large shift is already enough to completely fill the anisotropic $\sigma$ bands. We consider two effects on the $\sigma$ bands of a small upward shift of $E_F$. i) the band dispersion and therefore the Fermi velocity changes. At higher energies, particularly the in-plane Fermi velocity smaller,[34] corresponding to a smaller coherence length $\xi_{ab,\sigma} \propto v_F^{ab,\sigma}/\Delta_\sigma$ and ultimately a higher $H_{c2}$ and, since the out-of-plane Fermi velocity changes much less than the in-plane one, a smaller anisotropy $\gamma_H(0)$.[8] However, while the changes in the $\sigma$ band Fermi velocity upon a change of $E_F$ corresponding to 6.3% C substitution should lead to an increase of $H_{c2}^{\|c}(0)$ even taking into account the $T_c$ suppression, the effect is much smaller than the measured increase. The lower measured anisotropy $\gamma_H(0)$ is explained in a good part by the decrease of the anisotropy of $v_F^\sigma$ due to the Fermi level shift. The weaker $T$ dependence of $\gamma_H$ may be qualitatively understood as well, in that the $\pi$ band Fermi velocity, contributing at higher $T$,[9] does not change appreciably. ii) the shift weakens the superconductivity as evidenced by the decrease in $T_c$. This competes with the first effect, decreasing the magnitude of $H_{c2}$, but does not affect the $H_{c2}$ anisotropy.

These considerations show that *intrinsic changes* of the Fermi surface are important and *cannot be neglected* in the assessment of the effect of C substitution. They also show, however, that some properties, particularly the *strong increase of the magnitude of $H_{c2}$, cannot be accounted for by intrinsic changes*. Therefore we have to exclude both sets of bands being in the clean limit. Together with the previous conclusion that $\pi$ and $\sigma$ bands cannot both be in the dirty limit, it follows that either the σ bands are in the dirty limit and the π bands in the dirty limit, or the other way around. The former is unlikely for the following reasons: i) in pure MgB$_2$ crystals the π bands seem to be more dirty than the σ bands[32] ii) if the increase of σ intra-band scattering was the dominating effect of C substitution, a more pronounced decrease of $γ_H$ with increasing $T$ would be expected, as noted above. Our results thus suggest that the C substitution mainly drives the π bands dirtier, making the superconductivity within the π bands more robust with respect to an external field. The upper critical field in effect is no longer determined by the σ bands alone, even at low temperature. This effect probably contributes to the decrease of $γ_H$(0). A partial substitution within the B layers affecting mainly the π bands instead of the σ bands may seem counter-intuitive at first. However, since the π bands are already rather dirty in pure MgB$_2$ crystals, it is reasonable to expect the effective coherence length in the π bands to be more sensitive to increased scattering. We note that a similar conclusion of scattering being increased mainly in the π bands was drawn from results on MgB$_2$ thin films containing carbon.[35] However, the films in question have a much less well defined defect structure than the single crystal studied in the current work.

The three main effects of C substitution affecting $H_{c2}$ and its anisotropy can therefore be tentatively assigned to 1) dirtier π bands as a result of scattering, 2) lowered Fermi velocities within the σ bands due to the electron doping, 3) weakening of the superconductivity due to the decreased density of states in the σ bands upon electron doping. The first two effects lead to an increase of $H_{c2}$, whereas the third effect leads to a decrease of $H_{c2}$ and eventually the destruction of superconductivity altogether. Combining the present data with those of Refs. 14, 15, and 16, the result seems to be a quite strong initial increase of $H_{c2}$ up to roughly 3% to 5% of carbon, followed by a substitution range where $H_{c2}$ does not vary too much, presumably because the three effects considered above cancel there (see Fig. 5). In the figure, we did not include the data of Ref. 19, since Masui *et al.* estimated coherence lengths using WHH formula,[36] i.e. from the $dH_{c2}/dT$ slope near $T_c$, that is very indirect, and assumes standard-temperature dependence of $H_{c2}$ (which is not the case, as we know, and this is likely the origin of the discrepancy in $H_{c2}^{\|ab}$ for higher carbon contents). For higher substitution levels $H_{c2}$ decreases again until superconductivity is finally destroyed at about 12% as the σ bands completely fill. On the other hand, $\gamma_H$ decreases monotonically upon C substitution, reaching values close to 1 just before destruction of superconductivity.[15] In order to get a comprehensive picture, additional experimental data of the influence of Mg site substitution in $MgB_2$ on $H_{c2}$ and the PE are highly desirable.

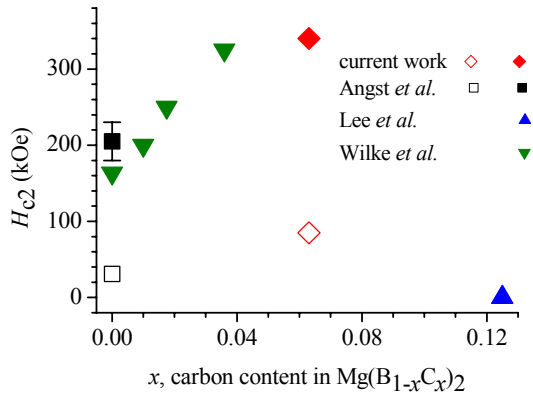

**FIG. 5. Upper critical field $H_{c2}$ of $Mg(B_{1-x}C_x)_2$ as a function of carbon content. Open symbols correspond to $H_{c2}^{\|c}(0)$ (squares – Ref. 4). Full symbols correspond to $H_{c2}^{\|ab}(0)$ (squares – Ref. 27, triangles up – Ref. 15, triangles down – Ref. 16).**

In conclusion, the influence of partial carbon substitution into the boron site of $MgB_2$ on the anisotropic upper critical field and peak effect can be explained as disorder effects only if assumed that the main influence of the disorder is to make the π bands more dirty. The mechanism by which C substitution (even in single crystals) may cause scattering mainly in the π bands is a mystery yet to be solved. In addition to disorder, intrinsic effects of the substitution on the Fermi surface have to be taken into account as well. The main intrinsic effect is to dope the system with additional electrons, shifting the Fermi energy $E_F$ upwards, leading to a decrease of the σ band dispersion at the Fermi level and a weakening of superconductivity. The competing effects of both intrinsic and extrinsic nature determine the evolution of the upper critical field magnitude (and anisotropy) with the carbon substitution, with a broad maximum below 400 kOe at substitution levels between about 4 and 7%.

We would like to thank V. P. Antropov for very valuable comments concerning the changes in the Fermi surface as a result of carbon substitution in $MgB_2$ as well as for the critical reading of the manuscript. We would also like to thank A. Gurevich for a critical reading and useful comments concerning the scattering discussion. This work was supported in part by the European Community (contract ICA1-CT-2000-70018), Swiss Office BBW (Nr. 02.0362), and by the Polish State Committee for Scientific Research (project 5 P03B 12421). Ames Laboratory is operated for the US Department of Energy (contract W-7405-Eng-82). MA thanks the Swiss National Science Foundation for financial support.

[1] J. Nagamatsu, N. Nakagawa, T. Murakanaka, Y. Zenitani, and J. Akimitsu, Nature (London) **410**, 63 (2001).

[2] For a recent review, see, P. C. Canfield and G. W. Crabtree, Physics Today **56** (3), 34 (2003).

[3] A. Y. Liu, I. I. Mazin, and J. Kortus, Phys. Rev. Lett. **87**, 087005 (2001); H. J. Choi, D. Roundy, H. Sun, M. L. Cohen, and S. G. Louie, Nature (London) **418**, 758 (2002).

[4] M. Angst, R. Puzniak, A. Wisniewski, J. Jun, S. M. Kazakov, J. Karpinski, J. Roos, and H. Keller, Phys. Rev. Lett. **88**, 167004 (2002).

[5] S. L. Bud'ko and P. C. Canfield, Phys. Rev. B **65**, 212501 (2002); L. Lyard, P. Samuely, P. Szabo, T. Klein, C. Marcenat, L. Paulius, K. H. P. Kim, C. U. Jung, H.-S. Lee, B. Kang, S. Choi, S.-I. Lee, J. Marcus, S. Blanchard, A. G. M. Jansen, U. Welp, G. Karapetrov, and W. K. Kwok, Phys. Rev. B **66**, 180502(R) (2002); U. Welp, A. Rydh, G. Karapetrov, W. K. Kwok, G. W. Crabtree, Ch. Marcenat, L. Paulius, T. Klein, J. Marcus, K. H. P. Kim, C. U. Jung, H.-S. Lee, B. Kang, and S.-I. Lee, Phys. Rev. B **67**, 012505 (2003); L. Lyard, P. Szabo, T. Klein, J. Marcus, C. Marcenat, K. H. Kim, B. W. Kang, H. S. Lee, and S. I. Lee, Phys. Rev. Lett. **92**, 057001 (2004).

[6] M. Angst, D. Di Castro, R. Puzniak, A. Wisniewski, J. Jun, S. M. Kazakov, J. Karpinski, S. Kohout, H. Keller, Physica C, in print; cond-mat/0304400.

[7] A. Rydh, U. Welp, A. E. Koshelev, W. K. Kwok, G. W. Crabtree, R. Brusetti, L. Lyard, T. Klein, C. Marcenat, B. Kang, K. H. Kim, K. H. P. Kim, H.-S. Lee, and S.-I. Lee, cond-mat/0308319.

[8] For a review, see, M. Angst and R. Puzniak, in 'Focus on Superconductivity' edited by B. P. Martines (Nova Science Publishers, New York, 2004), pp. 1-49; cond-mat/0305048.

[9] P. Miranovic, K. Machida, and V. G. Kogan, J. Phys. Soc. Jpn. **72**, 221 (2003).

[10] A. Gurevich, Phys. Rev. B **67**, 184515 (2003).

[11] T Dahm and N. Schopohl, Phys. Rev. Lett. **91**, 017001 (2003); A. A. Golubov and A. E. Koshelev, Phys. Rev. B **68**, 104503 (2003); A. E. Koshelev and A. A. Golubov, Phys. Rev. Lett. **92**, 107008 (2004).

[12] D. Larbalestier, A. Gurevich, D. M. Feldmann, and A. Polyanskii, Nature (London) **414**, 368 (2001).

[13] For brief reviews of substitutional chemistry on $MgB_2$ see, for example, R. J. Cava, H. W. Zandbergen, and K. Inumarua, Physica C **385**, 8 (2003); R. A. Ribeiro, S. L. Bud'ko, C. Petrovic, and P. C. Canfield, ibid. **385**, 16 (2003).

[14] R. A. Ribeiro, S. L. Bud`ko, C. Petrovic, and P. C. Canfield, Physica C **384**, 227 (2003).

[15] S. Lee, T. Masui, A. Yamamoto, H. Uchiyama, and S. Tajima, Physica C **397**, 7 (2003).

[16] R. H. T. Wilke, S. L. Bud'ko, P. C. Canfield, D. K. Finnemore, R. J. Suplinskas, and S. T. Hannahs, cond-mat/0312235.

[17] E. Ohmichi, T. Masui, S. Lee, S. Tajima, and T. Osada, cond-mat/0312348.

[18] M. Pissas, D. Stamopoulos, S. Lee, and S. Tajima, cond-mat/0312350.

[19] T. Masui, S. Lee, and S. Tajima, cond-mat/0312458.

[20] P. Samuely, Z. Holanova, P. Szabo, J. Kacmarcik, R. A. Ribeiro, S. L. Bud'ko, and P. C. Canfield, Phys. Rev. B **68**, 020505 (2003); H. Schmidt, K. E. Gray, D. G. Hinks, J. F. Zasadzinski, M. Avdeev, J. D. Jorgensen, and J. C. Burley, Phys. Rev. B **68**, 060508(R) (2003).

[21] J. Karpinski, S. M. Kazakov, J. Jun, M. Angst, R. Puzniak, A. Wisniewski, and P. Bordet, Physica C **385**, 42 (2003).

[22] M. Avdeev, J. D. Jorgensen, R. A. Ribeiro, S. L. Bud'ko, and P. C. Canfield, Physica C **387**, 301 (2003).

[23] S. M. Kazakov, R. Puzniak, K. Rogacki, A. V. Mironov, N. D. Zhigadlo, J. Jun, Ch. Soltmann, B. Batlogg, and J. Karpinski, in preparation.

[24] I. I. Mazin, O. K. Andersen, O. Jepsen, O. V. Dolgov, J. Kortus, A. A. Golubov, A. B. Kuz'menko, and D. van der Marel, Phys. Rev. Lett. **89**, 107002 (2002).

[25] A. V. Sologubenko, J. Jun, S. M. Kazakov, J. Karpinski, and H. R. Ott, Phys. Rev. B **65**, 180505(R) (2002).

[26] M. Angst, R. Puzniak, A. Wisniewski, J. Jun, S. M. Kazakov, and J. Karpinski, Phys. Rev. B **67**, 012502 (2003).

[27] M. Angst, R. Puzniak, A. Wisniewski, J. Roos, H. Keller, P. Miranovic, J. Jun, S. M. Kazakov, and J. Karpinski, Physica C **385**, 143 (2003).

[28] J. Karpinski, M Angst, J. Jun, S. M. Kazakov, R. Puzniak, A. Wisniewski, J. Roos, H. Keller, A. Perucchi, L. Degiorgi, M. R. Eskildsen, P. Bordet, L. Vinnikov, and A. Mironov, Supercond. Sci. Technol. **16**, 221 (2003).

[29] D. R. Tilley, Proc. Phys. Soc. London **86**, 289 (1965).

[30] No structural evidence for such a scenario was obtained, however. Possible origins of the unusual peak effect behavior in substituted $MgB_2$ will be further considered elsewhere.

[31] It is difficult to assign the increase in the residual resistivity to a particular of these scattering rates without assumptions on effect of C substitution on the electronic structure, but with assumption that the main conductance channel remains the $\pi$ bands as in unsubstituted $MgB_2$ the drastic increase in residual resistivity would seem to imply increased intraband scattering in the $\pi$ bands.

[32] A. Carrington, P. J. Meeson, J. R. Cooper, L. Balicas, N. E. Hussey, E. A. Yelland, S. Lee, A. Yamamoto, and S. Tajima, Phys. Rev. Lett. **91**, 037003 (2003).

[33] See, e.g., F. Bouquet, Y. Wang, I. Sheikin, T. Plackowski, A. Junod, S. Lee, and S. Tajima, Phys. Rev. Lett. **89**, 257001 (2002).

[34] I. I. Mazin and V. P. Antropov, Physica C **385**, 49 (2003).

[35] A. Gurevich, S. Patnaik, V. Braccini, K. H. Kim, C. Mielke, X. Song, L. D. Cooley, S. D. Bu, D. M. Kim, J. H. Choi, L. J. Belenky, J. Giencke, M. K. Lee, W. Tian, X. Q. Pan, A. Siri, E. E. Hellstrom, C. B. Eom, and D. C. Larbalestier, Supercond. Sci. Technol. **17**, 278 (2004).

[36] N. R. Werthamer, N. Helfand, and P. C. Hohenberg, Phys. Rev. **147**, 295 (1966).